\documentclass[conference]{IEEEtran}
\usepackage{cite}
\usepackage{amsmath,amssymb,amsfonts}
\usepackage{algorithmic}
\usepackage{graphicx}
\usepackage{textcomp}
\usepackage{xcolor}
\usepackage{algorithm}
\usepackage{algorithmic}
\usepackage{multirow}
\usepackage{multicol}
\def\BibTeX{{\rm B\kern-.05em{\sc i\kern-.025em b}\kern-.08em
    T\kern-.1667em\lower.7ex\hbox{E}\kern-.125emX}}
\begin{document}

\title{Secure and Decentralized Swarm Behavior with Autonomous Agents for Smart Cities}

\author{\IEEEauthorblockN{Rafer Cooley}
\IEEEauthorblockA{Computer Science Department\\
University of Wyoming\\
Laramie, WY\\
rcooley2@uwyo.edu}
\and
\IEEEauthorblockN{Shaya Wolf}
\IEEEauthorblockA{Computer Science Department\\
University of Wyoming\\
Laramie, WY\\
swolf4@uwyo.edu}
\and
\IEEEauthorblockN{Mike Borowczak}
\IEEEauthorblockA{Computer Science Department\\
University of Wyoming\\
Laramie, WY\\
mike.borowczak@uwyo.edu}}


\maketitle

\begin{abstract}
Unmanned Aerial Vehicles (UAVs), referenced as drones, have advanced to consumer adoption for hobby and business use. Drone applications, such as infrastructure technology, security mechanisms, and resource delivery, are just the starting point. More complex tasks are possible through the use of UAV swarms. These tasks increase the potential impacts that drones will have on smart cities, modern cities which have fully adopted technology in order to enhance daily operations as well as the welfare of it's citizens. Smart cities not only consist of static mesh networks of sensors, but can contain dynamic aspects as well including both ground and air based autonomous vehicles. 



Networked computational devices require paramount security to ensure the safety of a city. To accomplish such high levels of security, services rely on secure-by-design protocols, impervious to security threats. Given the large number of sensors, autonomous vehicles, and other advancements, smart cities necessitates this level of security. The SHARK protocol (Secure, Heterogeneous, Autonomous, and Rotational Knowledge for Swarms)  ensures this kind of security by allowing for new applications for UAV swarm technology. Enabling drones to circle a target without a centralized control or selecting lead agents, the SHARKS protocol performs organized movement among agents without creating a central point for attackers to target. Through comparisons on the stability of the protocol in different settings, experiments demonstrate the efficiency and capacity of the SHARKS protocol.
	
\end{abstract}



\section{Introduction}

Smart cities incorporate cutting-edge technology into the infrastructure of the city to simplify and enhance daily operations. 
These technologies can take many forms, including technology from self-driving vehicles to vast networks of sensors. As drone research progresses, their benefit to a smart city expands. Although singular drones have much to offer, their combined work can carry out much more intricate plans. Through simple individual behaviors, swarms complete complex tasks necessary in a smart city.  UAV (Unmanned Ariel Vehicles) swarms benefit smart cities in many ways. Buildings, bridges, and roads require maintenance that is simple for a drone but dangerous for a human being. UAVs locating injured citizens quickly during an emergency situation implicates a much lower mortality rate. Furthering these benefits, the SHARKS protocol (Secure, Heterogeneous, Autonomous, and Rotational Knowledge for Swarms) allows agents to circle a target with only two simple rules. This allows for increased swarm functionality with little additional computation power. 

Most advanced sensor technologies benefit from autonomy, meaning that once a user has installed and configured the hardware, the technology runs itself. After configuration, it is up to the machine to perform the proper tasks assigned to it without fail. Ideally, UAV swarms are self-deploying and can stabilize from many different initializations. Further, the swarms are decentralized, meaning that there are no central nodes directing traffic or any lead nodes for the rest to follow. This enhances the security of swarms with autonomous decentralized protocols such as SHARKS.

The SHARKS protocol ensures movement without collision and aims to get all of the agents in a specific range of the target. By doing so, these swarms maintain stable rotations without collisions. This stability relies on the efficiency of the swarm as well as the carrying capacity of the specified range from the target. The efficiency of the swarm depends on the population size, the initial distribution of the agents, and the distance moved each epoch. The carrying capacity depends on the population size, the ideal distance from the target, and the acceptable range from the ideal distance.

\section{Applications}



\subsection*{Swarm Behavior}
Swarm robotics provides a place for simple algorithms to be applied on complex scales. Similar to the way human beings perform individual tasks to contribute to an intricate metropolis, agents can also contribute to swarm interests through interplay between discrete goals. Robotic swarms have the potential to transform how we complete tasks from parcel delivery to tending to rooftop gardens. Previous work shows that robotic swarms can act like flocks of birds or schools of fish through decentralized algorithms \cite{algoForSwarmPaper}. Although this flocking behavior enables swarms to move and congregate, sometimes more specific movement is necessary. Rule sets can be surprisingly simple. With a smaller rule set and less information to store in any given agent, SHARKS provides new movement behaviors beyond flocking. Given only two elementary objectives, agents circle a target, extending the possibilities for swarm technology. With this behavior, small UAV swarms can tend to a rooftop garden by collecting data from soil sensors, protecting the premises from unwelcome animals, and monitoring environmental variables such as air quality. While most smart cities employ static sensor networks in applications such as these, they are more vulnerable to attack then a network with dynamic agents, like UAV swarms. By sharing information through moving drones, sensor networks can scatter their information while still keeping it organized and useful. This enhances the security of sensor networks as well as expanding their potential. 

\subsection*{UAV Cooperation}
Swarm technologies inherently need agent cooperation built into the system in order for the swarm to be useful in any context. UAVs are much more powerful in a cooperative swarm than they are alone. Due to their small size and low computational power, an individual drone has limited functionality and impact. However, by cooperating with other UAVs, much more complex tasks can be completed. Research has flourished in this area, providing autonomous cooperation to maximize information obtained from sensors and enhance the potential tasks of these swarms \cite{coopUAVPaper}. Allowing swarms to find a target on the ground, current research aids UAVs in broader goals such as target acquisition and tracking. The SHARKS protocol expands these applications by facilitating cooperative movements to encircle a target. Drones fit with water delivery methods could extinguish a fire, drones fit with parcel delivery methods could deliver necessary goods to areas that would be too difficult to get to on the ground, and UAVs fit with appropriate sensors could detect and form a perimeter around environmental threats such as oil spills, forest fires, and nuclear material sites. Drone swarms are also more fit to clean up after a natural disaster or deliver water to a remote village. For disaster clean-up or long-distance resource delivery, it makes more sense to utilize cooperative artificial robotic agents than to risk to human lives.

\subsection*{Risky or Hazardous Situations}
The SHARKS protocol can significantly reduce the risk of current hazardous jobs. First responders enter burning buildings, attend to medical emergencies in potentially dangerous situations, and thrive in many other unpredictable and threatening settings. By outsourcing dangerous tasks to UAV swarms, smart cities strive for higher safety for its civilians. Even simple reconnaissance into a dangerous situation can provide safer access points, information for swift goal-completion, and ideal exit strategies when human interaction is required. By circling emergency cites, drones can provide information to first responders such as where people may be in a burning building or finding people who need medical assistance after a large car crash. Additionally, human operators must scale structures that are several stories tall and difficult to navigate to perform simple maintenance tasks. By deploying robotic swarms for these tasks, routine maintenance can be provided without risk to human operators. Further, There are many other areas that are either difficult or dangerous for human beings to navigate. In these high-risk situations, dispatching a UAV swarm could not only save lives, but also allows for more efficient problem resolution. For example, allowing a UAV swarm to tend to critical infrastructure is not only safer for human beings, but also ensures a more efficient maintenance schedule and less catastrophes over time. Further, due to it's autonomous and decentralized nature, the SHARKS protocol enhances the security of such infrastructure. 

\subsection*{Targeting, Tasking, and Tracking}
Other robotic swarms have looked at targeting, tasking and tracking \cite{targetingPaper}. Also through a decentralized approach, these swarms can take-off, search, task, and track without selected leaders in the swarm. The SHARKS protocol adds further functionality with little computational cost to each agent. With the addition of SHARKS, these swarms can circle targets and track them in formation. This extends the applications of robotics swarms to allow for dynamic targets and smarter tracking services. This swarm behavior can aid in sensor repair in an urban area and thereby ensure more efficient resource use. Further, a decentralized approach to surveillance mechanisms through UAV swarms provides privacy for systems that are currently open to abuse from not only criminals, but also from within the institution these systems are meant to benefit. An autonomous approach protects from many other privacy issues, as robotic swarms would operate apart from the extra objectives that human operators may have for such technology.

\subsection*{Adversarial Swarms}
Research has lead to robotic swarm applications in navigating adversarial environments. UAVs can maneuver dangerous environments and no-fly zones, allowing for their use in instances where it would be too dangerous to send a human being \cite{adversarialPaper}. Further, the adversarial landscape could potentially include enemy UAV swarms. Organized crime groups have taken advantage of advances in consumer drone technology and use this technology to circumvent law enforcement efforts. Smart cities ensure security in an advanced environment by using drone technology to interdict enemy drones \cite{interdictPaper}. However, current UAV technology relies on using various signals for communication and maneuvering in the field. This allows for GPS spoofing, a large vulnerability in current technology \cite{spoofing}.
Due to these vulnerabilities, smart cities deploying drones within their infrastructure face a potential for crime groups to tamper with or intercept the drones for use in illicit activities. To prevent this, a secure-by-design implementation of drone protocols is crucial. Since the SHARKS protocol does not rely on conventional positioning services, it is resilient to GPS jamming techniques thus negating the attack vector and increasing the difficulty for criminal groups to overtake the drones.



\section{Motivation}
\subsection*{Security}
As vulnerability landscapes evolve and cities add technology to their infrastructure, security systems strive to beat the curve. Services essential to daily operations become reliant on secure and dependable protocols for implementation. The increasing pace of technology is followed closely by the sophistication of attacks. Current development trends promote technology using existing outdated technologies and as a result companies have failed to solve issues in securing new technology. 
For example, drone technology currently relies upon traditional GPS services to provide locational awareness. However, GPS can easily be jammed leading to incorrect data and potential crashes. As new drone technology surfaces, all of the protocols that rely on GPS locations simply recycle the same security flaws. 
With the SHARKS protocol, locational awareness becomes relative. Through a decentralized movement algorithm, agents within a swarm can circle a target with only the location of the target and the location of the nearest neighboring agent. Because the target provides the desired location that affects the movements for the entire swarm, it is not reliant on GPS and therefore not vulnerable to GPS security flaws, making it a viable option for future drone security. Current research in the field pertains to decentralized movement of self deployed swarms \cite{geometricPatternsPaper}. The SHARKS protocol builds on this research to gain new behaviors from self-deployed agents in multiple different settings without the need to layer existing security technologies on top of the new behavior. 

\subsection*{Resiliency}

New UAV swarm technology strives for secure-by-design systems that are resilient to adversarial interests. Much like we have seen in secured cryptocurrencies, decentralized systems are ideal for multi-agent tasks. Due to the lack of a central server or selected agent leaders to direct the behavior of the group, the SHARKS protocol is impervious to adversaries and continues to function even if agents are lost. Integrity in the system is ensured through not allowing any one drone to carry too much information gathered by the swarm or hold too much influence over the behavior of the swarm. Each agent only knows the information needed to perform its individual behaviors and carries out those behaviors.

\subsection*{Obstacle Avoidance}
There are often obstacles in a terrain that challenge a swarms ability to adapt to it's surroundings. Current research is interested in not only avoiding these obstacles, but also other agents in a swarm \cite{obstaclesPaper}; they must maneuver in the same physical space as each other without colliding. Due to the security benefits of a decentralized system and built-in obstacle avoidance, the SHARKS protocol allows for agents to navigate in an area without colliding with each other, while only knowing the location of the target and their nearest neighboring agent.

\section{SHARKS Protocol}
\subsection*{Algorithm}
The SHARKS protocol relies on two simple rules that each agent must follow to maintain target-circling behavior. Each agent must move towards the target (Center Rule) and away from their nearest neighbor (Dispersion Rule). This is done with a certain degree of rotation ($r$) to ensure the agents are moving at a reasonable speed around the target. 

\subsubsection{Center Rule}
Each agent aims to move to a specified distance ($\delta$) from the target. By doing this, each agent stays within a specific distance range from the target ($\delta \pm \epsilon$). This rule is accomplished through the following algorithm:

\begin{algorithm}[H]
\caption{Center Rule Algorithm}
\begin{algorithmic}[1]
  \IF {($\delta - dist > \epsilon$)}
    \IF {($dist - c = empty$)}
      \STATE move backwards $c$ units
    \ENDIF
  \ENDIF
  \IF {($\delta - dist < -\epsilon$)}
    \IF{($dist + c = empty$)}
      \STATE move forwards $c$ units
    \ENDIF
  \ENDIF
\end{algorithmic}
\end{algorithm}

In this algorithm, $dist$ is the distance that agent is away from the target and $c$ is the number of units each agent can move in one iteration of the algorithm to satisfy the center rule. Note that $empty$ implies that there are no agents/obstacles at a given location. This allows agents to move within $\delta \pm \epsilon$ units from the target. 

\subsubsection{Dispersion Rule}
Each agent aims to move away from their nearest neighbor and they do this at an clockwise angle of $(180+r)^{\circ}$. By doing this, each agent distances itself from every other and maintains a rotation around the target. This rule is accomplished through the following algorithm:

\begin{algorithm}[H]
\caption{Dispersion Rule Algorithm}
\begin{algorithmic}[1]
  \STATE Determine heading of nearest neighbor
  \STATE Rotate heading clockwise by 180 + $r$ degrees
  \IF {($dist + d = empty$)}
    \STATE move $d$ units
  \ENDIF
\end{algorithmic}
\end{algorithm}

\begin{figure}
\includegraphics[width=\linewidth]{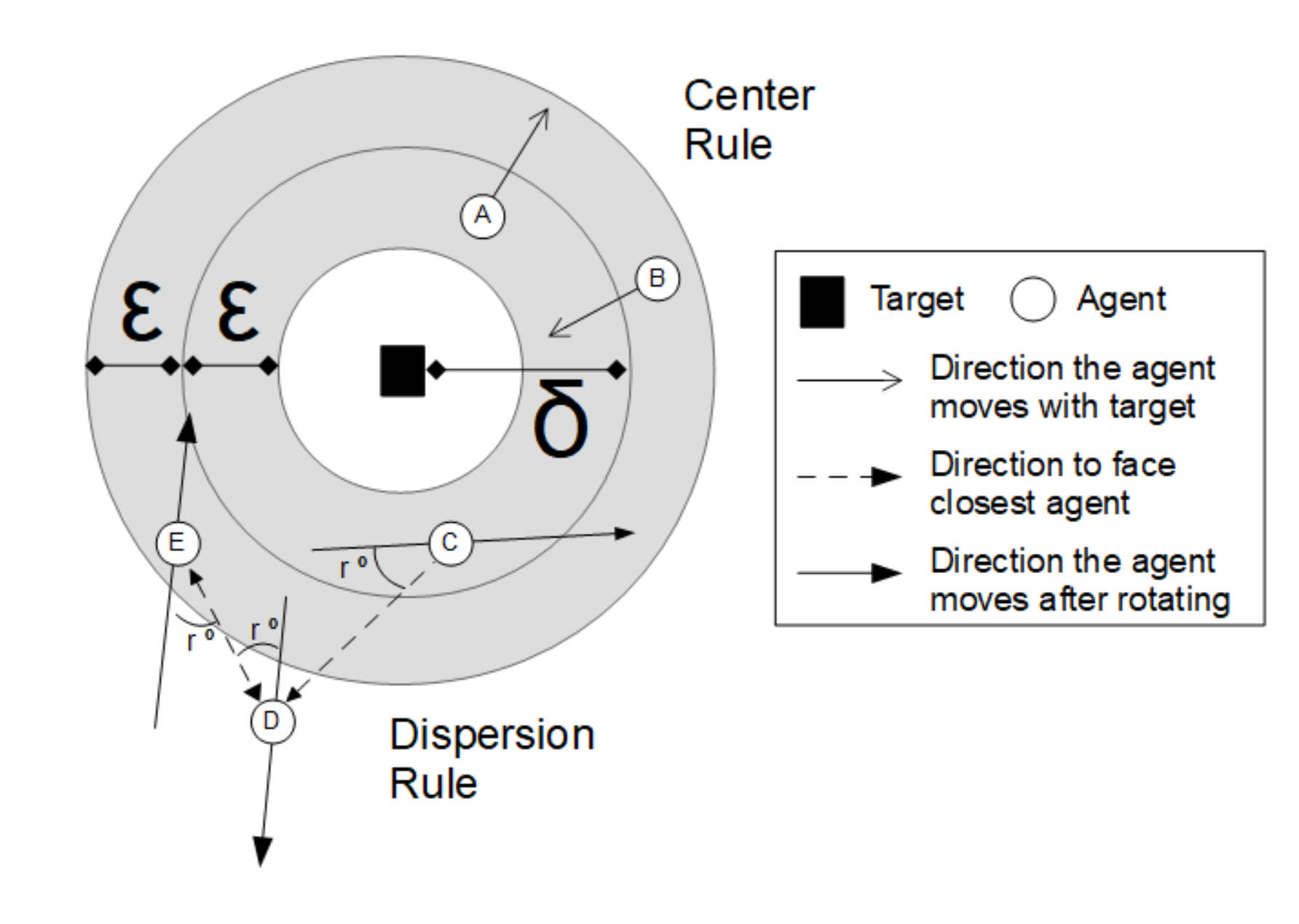}
\caption{The two main rules for agent movement in SHARKS visualized. While not pictured, agents A and B also have a dispersion rule applied prior to movement. Similarly, agents C,D, and E have a center rule applied prior to movement.}
\label{shark-diagram}
\end{figure}

In this algorithm, $dist$ is the distance the agent is away from the target and $d$ is the number of units each agent can move in one iteration of the algorithm to satisfy the dispersion rule. Again, note that $empty$ implies that there are no agents at a given location. This allows agents to distance themselves from one another and rotate around the target. These rules can be seen in Figure \ref{shark-diagram}.

Since each agent can move $c$ units to satisfy the Center Rule, and $d$ units to satisfy the Dispersion Rule, they may find it more beneficial if the $c \neq d$. In other words, given the ability to favor one rule over the other, a swarm may reach stability quicker than if they aim to satisfy each rule equally. Therefore, these experiments look at two separate ratios of these values. The first $d:c$ ratio is simply 1:1, where each agent moves equal distances to satisfy each rule. The second $d:c$ ratio is 3:4, where for every three units each agent moves to satisfy the Dispersion Rule, they move four units to satisfy the Center Rule.

\subsection*{Initial Distribution}
There are a number of distributions that the agents can be initialized at, such that swarms can be deployed in numerous layouts and still circle the target. Agents can be initialized randomly, boxed, linearly, or radially linearly. Regardless of the initial distribution, two or more agents cannot be initialized in the exact same space on the field. 
\begin{itemize}
\item \textbf{Random} - With a random initialization, agents are scattered randomly across the field. For this distribution, some agents may or may not start within the target distance from the target. 
\item \textbf{Boxed} - With a boxed initialization, agents are randomly placed within a set of coordinates. For these experiments, the boxes each have a ten unit width and a ten unit height. They are placed in nine separate locations, in each corner of the landscape, on each side, and in the center of the landscape.
\item \textbf{Linear} - With a linear initialization, agents are randomly set along a horizontal line 20 units above the target and spanning out past the desired range.
\item \textbf{Collinear} - With a collinear initialization, agents are randomly set along a vertical line across the center of the target (therefore collinear with the target) and spanning out past the desired range. 
\end{itemize}

\section{Experiments}
\subsection*{Efficiency}
These experiments aim to find the number of epochs necessary for all of the agents in a swarm to fall into the desired area on the landscape. Once all of the agents are within the specified distance from the target, the swarm is stable. The SHARKS protocol is tested under many different variances including population size, rotation amount ($r$), initial distributions, and ratios between the center rule and dispersion rule ($c$ and $d$). This will measure how efficiently different kinds of swarms can organize into a stable equilibrium. We compare these results based on the epoch in which each swarm reaches stability. Each experiment is ran five times, with the exception of the box initial distribution. These experiments are only ran twice since there are nine different iterations with this distribution and some act similarly to others. For example, experiments initialized in a corner are run eight times, twice for each corner.

\subsection*{Capacity}
These experiments aim to find the ideal population size for the SHARKS protocol. This is tested under many different variables as well, including population size, ideal distance from the target ($\delta$), and allowable distance from the target distance ($\epsilon$). From these experiments, we can see the ideal population size of the SHARKS protocol in different situations with more or less room for the agents to operate. Each experiment is run five times and will measure the the carrying capacity of the SHARKS protocol. We again compare these results based on the epoch in which each swarm reaches stability.

\section{Results}
\subsection*{Efficiency}

\subsubsection*{Random}
\begin{figure}[h]
\includegraphics[width=\linewidth]{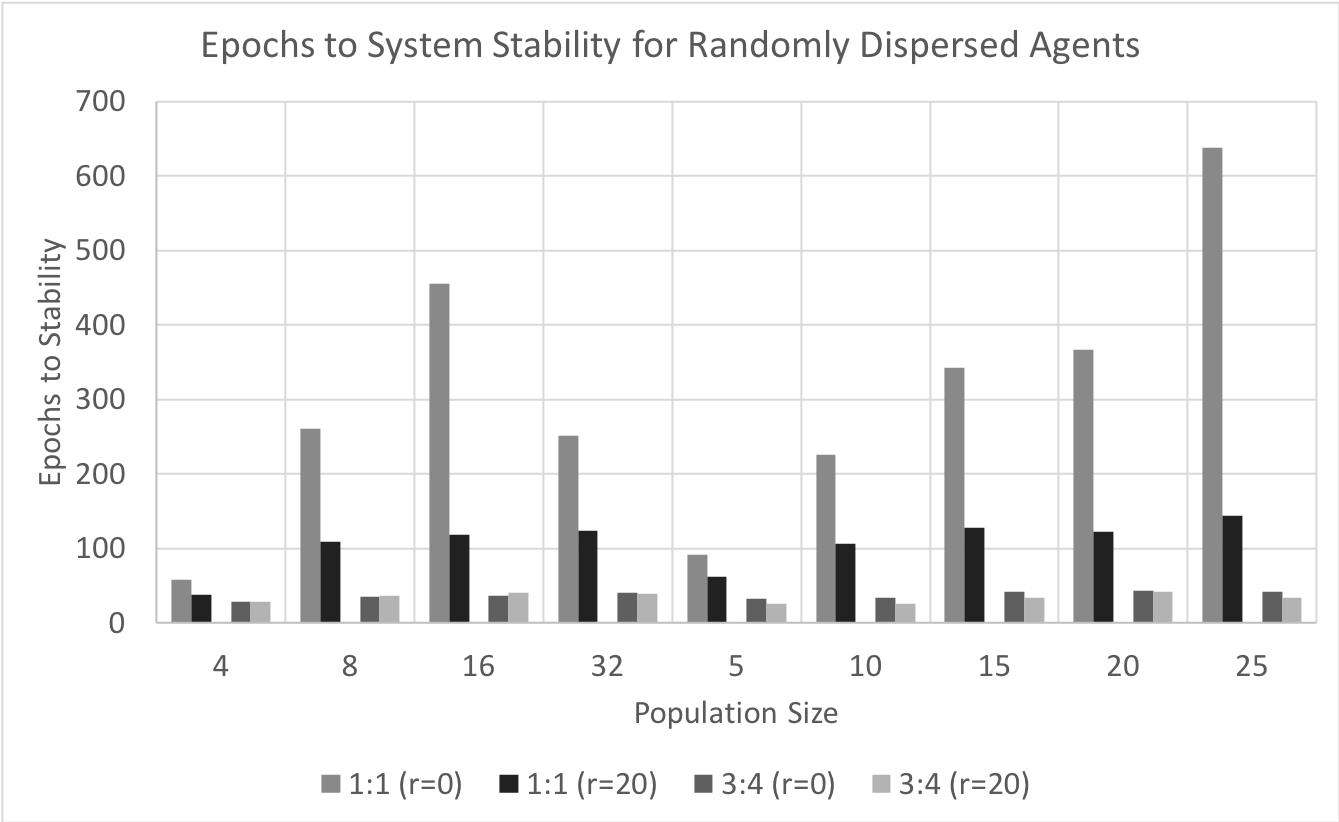}
\caption{Data from Table \ref{randomInit} plotted to show the impact of the c:d ratio as well as a non-zero rotation $r$ on the number of epochs to stability.}
\label{random-init-graph}
\end{figure}

In the random initializations, the swarm reached stability rather quickly, around 30-40 epochs regardless of the population size when using a $d:c$ ratio of 3:4. Recall that the $d:c$ ratio is the number of units an agent can move across to satisfy either the dispersion rule or the center rule. So a ratio of 3:4 means that for every 4 units an agent moves to satisfy the center rule, they can only move 3 units to satisfy the dispersion rule. By ensuring that the agents weight the center rule more than the dispersion rule, stability is reached sooner since the agents are slightly more focused on getting to the target. Otherwise, it takes longer to get within the ideal distance as the agents are too concerned with rotating and avoiding one another. In other words, the dispersion rule enables the agents to circle the target and avoid collisions, but it forces agents outside of the ideal distance of the target. These results can be seen in Table \ref{randomInit} and Figure \ref{random-init-graph}. With a random initialization, this 3:4 ratio has ideal results, while a 1:1 ratio performs poorly. In the 1:1 ratio, the number of epochs needed to reach stability increases drastically with the population size, requiring more that 600 epochs in a swarm with 25 agents. 

This demonstrates the efficiency of the SHARKS protocol with random initialization. Clearly, smaller populations stabilize faster than larger populations. In the population sizes that double each time (sizes of 4, 8, 16, and 32), the number of epochs it takes to stabilize the swarm does not double. In the population sizes that grow statically (sizes of 5, 10, 15, 20, 25), the number of epochs it takes to stabilize the swarm grows rather linearly. This indicates that population size doesn't have any exponential affect on the stability of the swarm. Further, having a rotation of 20 degrees helps the swarm reach stability much faster. In some cases, reaching stability is done two to six times faster when a rotation is added to the nodes. The ideal swarm would be smaller and have a rotation ($r$) of 20 degrees. This rotation isn't as vital when the $d:c$ ratio is 3:4, however, when the ratio is 1:1, the rotation amount is crucial.

\renewcommand{\arraystretch}{1.5}
\begin{table}[H]
\caption{Random Initialization Results}

\centering
\resizebox{\columnwidth}{!}{
\begin{tabular}{|c|c||c|c|c|c||c|c|c|c|c|}
\cline{3-11}
\multicolumn{1}{}{} & \multicolumn{1}{}{} & \multicolumn{9}{|c|}{Population Size} \\
\hline
$d:c$ & $r$ & 4 & 8 & 16 & 32 & 5 & 10 & 15 & 20 & 25 \\
\hline
\multirow{2}{*}{3:4} & 0 & 28.60 & 35.00 & 36.80 & 39.80 & 32.20 & 33.40 & 41.60 & 43.20 & 42.20 \\
& 20 & 28.80 & 35.80 & 39.80 & 38.40 & 26.20 & 26.20 & 34.20 & 41.20 & 33.80\\
\hline
\multirow{2}{*}{1:1} & 0 & 58.20 & 260.00 & 455.40 & 453.60 & 92.00 & 226.00 & 342.80 & 366.60 & 638.40\\
& 20 & 37.80 & 108.60 & 118.60 & 123.40 & 61.40 & 106.80 & 128.00 & 122.60 &144.20\\
\hline
\end{tabular}
}
\vspace{2mm}
\label{randomInit}
\end{table}

\subsubsection*{Boxed}
For experiments with a boxed initialization, all of the agents are initialized in a specified box on the field. For these experiments, the agents are initialized in 10 by 10 unit boxes. These boxes are placed in the corners, centered along the sides, top, and bottom, and placed in the center on top of the target. This allowed for a variety of results. Among these initializations, the populations initialized in a corner took the longest time to stabilize. Compared to random and linear initializations, cornered boxes performed better than random placements and slightly worse than linear placements. The boxes that were places along the sides performed about the same as those placed on the top or the bottom of the field. These experiments stabilized faster than both random and linear initializations. The best results were seen in the boxes placed on the center of the field, taking only 26 epochs in the worst case. The same experiment initialized randomly took 454 epochs on average to stabilize, took 169 epochs on average to stabilize with a linear initialization, and wasn't able to stabilize at all in a collinear initialization.

Unlike the other experiments, boxed initializations saw more resiliency towards adjustments made to the rotation amount. Where collinear initializations couldn't even reach stability without a rotation amount, the boxed experiments saw little effect on stability without a rotation amount. In fact, in the case were the boxes were initialized on the center of the field, small populations benefited from lacking a rotation amount, stabilizing faster for population sizes less than 12. Similar to the other experiments however, boxed initializations always performed better with a $d:c$ ratio of 3:4 rather than a 1:1 ratio. This can be seen across all population sizes, box placements, and rotation amounts. All of these results can be seen in Table \ref{boxInit}.

\renewcommand{\arraystretch}{1.5}

\begin{table}[]
\caption{Boxed Initialization Results}
\centering
\resizebox{\columnwidth}{!}{
\begin{tabular}{|c|c|c||c|c|c|c||c|c|c|c|c|}
\cline{4-12}
\multicolumn{1}{}{} & \multicolumn{1}{}{} & \multicolumn{1}{}{} & \multicolumn{9}{|c|}{Population Size}\\
\hline
$Box$ & $d:c$ & $r$ & 4 & 8 & 16 & 32 & 5 & 10 & 15 & 20 & 25\\
\hline
\multirow{4}{*}{\rotatebox[origin=c]{90}{Corners}} & \multirow{2}{*}{3:4} & 0 & 29.0 & 36.0 & 34.4 & 46.0 & 36.2 & 39.0 & 32.8 & 41.2 & 38.2\\
 & & 20 & 26.6 & 37.2 & 35.2 & 41.6 & 33.8 & 31.6 & 33.8 & 36.2 & 36.8\\
\cline{2-12}
 & \multirow{2}{*}{1:1} & 0 & 67.0 & 62.6 & 248.8 & 208.6 & 80.8 & 75.6 & 177.4 & 179.4 & 154.6\\
 & & 20 & 48.2 & 76.0 & 81.2 & 103.0 & 48.8 & 61.0 & 73.8 & 78.8 & 89.6\\
\hline
\multirow{4}{*}{\rotatebox[origin=c]{90}{Sides}}& \multirow{2}{*}{3:4} & 0 & 15.8 & 17.4 & 23.0 & 26.6 & 17.0 & 19.8 & 19.8 & 22.0 & 22.0\\
 & & 20 & 14.6 & 15.2 & 19.6 & 26.2 & 16.0 & 18.8 & 17.6 & 23.8 & 24.2\\
\cline{2-12}
 & \multirow{2}{*}{1:1} & 0 & 35.6 & 47.8 & 53.2 & 52.4 & 43.4 & 59.2 & 63.0 & 63.8 & 69.4\\
 & & 20 & 21.0 & 34.6 & 41.0 & 38.6 & 30.4 & 40.8 & 30.8 & 56.6 & 51.8\\
\hline
\multirow{4}{*}{\rotatebox[origin=c]{90}{Top/Bottom}} & \multirow{2}{*}{3:4} & 0 & 16.2 & 18.8 & 21.8 & 28.4 & 17.0 & 18.4 & 22.0 & 21.8 & 26.0\\
 & & 20 & 13.6 & 17.2 & 19.8 & 23.2 & 15.2 & 19.0 & 17.4 & 20.4 & 22.2\\
\cline{2-12}
 & \multirow{2}{*}{1:1} & 0 & 29.0 & 31.2 & 71.4 & 78.0 & 32.8 & 34.2 & 61.0 & 60.2 & 75.6\\
 & & 20 & 24.8 & 31.8 & 45.2 & 45.2 & 28.2 & 49.2 & 36.2 & 38.6 & 56.6\\
\hline
\multirow{4}{*}{\rotatebox[origin=c]{90}{Center}} & \multirow{2}{*}{3:4} & 0 & 6.4 & 8.0 & 13.6 & 14.8 & 6.4 & 9.2 & 12.4 & 12.4 & 14.4\\
 & & 20 & 7.4 & 8.0 & 12.4 & 14.6 & 7.8 & 10.8 & 11.8 & 12.4 & 12.0\\
\cline{2-12}
 & \multirow{2}{*}{1:1} & 0 & 5.6 & 12.4 & 18.4 & 26.6 & 5.6 & 12.0 & 17.4 & 19.0 & 18.0\\
 & & 20 & 6.6 & 16.0 & 17.6 & 21.4 & 9.8 & 12.2 & 18.4 & 19.8 & 16.0\\
\hline
\end{tabular}
}
\vspace{2mm}
\label{boxInit}
\end{table}

\subsubsection*{Linear}
Similar to the random initialization, linear initialization sees much better results using a 3:4 $d:c$ ratio than 1:1. When each agent moves an equal distance to satisfy each rule, they can't reach stability as quickly because they are distancing themselves from each other, pushing away from the target during the earlier epochs. Therefore, we can see that for linear initializations, a $d:c$ ratio of 3:4 is ideal. Also similar to random initializations, setting the rotation amount to 0 degrees versus 20 degrees doesn't affect the efficiency of the swarm with a 3:4 ratio however, it drastically affects the efficiency of a swarm with a 1:1 ratio. With a swarm size of 25, a 20 degree rotation allows the swarm to stabilize in over 300 epochs sooner. This shows us that a rotation amount isn't as crucial when the swarms utilize an appropriate $d:c$ ratio. Regardless, adding a rotation amount produces more target circling than the cases without a rotation amount. Although the agents still circle the target, without a rotation amount, this behavior is weaker than when a rotation amount of 20 degrees is employed. 

We see better results in the linear initialization than in the random initialization for either ratio and either rotation amount. Because the agents start closer to each other and have a good chance of starting closer to the target, the two rules work rather quickly to get the agents into formation. These results can be seen in Table \ref{linearInit}. Although linear initialization experiments outperformed random initializations, they did not perform better than boxed experiments. The most comparable initialization in the boxed experiments to the linear experiments is the cornered setups, which were the worst performing of the boxes experiments. Even so, they outperformed linear initializations for small populations and $d:c$ ratios of 1:1. 

\renewcommand{\arraystretch}{1.5}
\begin{table}[H]
\caption{Linear Initialization Results}
\centering
\resizebox{\columnwidth}{!}{
\begin{tabular}{|c|c||c|c|c|c||c|c|c|c|c|}
\cline{3-11}
\multicolumn{1}{}{} & \multicolumn{1}{}{} & \multicolumn{9}{|c|}{Population Size}\\
\hline
$d:c$ & $r$ & 4 & 8 & 16 & 32 & 5 & 10 & 15 & 20 & 25 \\
\hline
\multirow{2}{*}{3:4} & 0 & 21.60 & 32.75 & 32.25 & 37.75 & 35.25 & 31.50 & 32.75 & 34.00 & 34.25\\
& 20 & 26.75 & 28.00 & 30.00 & 33.00 & 28.50 & 32.50 & 24.50 & 34.50 & 31.50\\
\hline
\multirow{2}{*}{1:1} & 0 & 100.25 & 152.50 & 345.00 & 168.75 & 68.75 & 160.00 & 144.75 & 294.75  & 404.75 \\
& 20 & 44.50 & 97.75 & 94.25 & 48.00 & 54.50 & 73.50 & 102.25 & 112.50 & 79.75 \\
\hline
\end{tabular}
}
\vspace{2mm}
\label{linearInit}
\end{table}

\subsubsection*{Collinear}
Collinear initialization experiments revealed surprising results. In many cases, swarms initialized on the vertical axis of the target are not able to stabilize. These cases are marked with a zero in Table \ref{rlinearInit}. We can see that the most crucial element in this initialization is a rotation amount. All of the experiments that were not able to reach stability had zero degrees of rotation. Since the agents can be initialized across the target and all in a line, they have to be able to disperse at an angle away from each other, otherwise they can't get out of the line while satisfying the center rule. With a rotation of zero, none of the agents can move from the starting line. However, adding just one degree of rotation allows the agents to move to stability. 

As with the other initializations, the $d:c$ ratio affects the efficiency of the swarm. A ratio of 3:4 outperforms a ratio of 1:1 for all population sizes. As with all of the initializations, the $d:c$ ratio isn't as influential as the rotation amount. In a collinear initialization, this fact is much more obvious. In the larger population sizes, an appropriate $d:c$ ratio allowed the swarms to stabilize almost four time faster. Further, the population sizes, similar to the other initializations, saw a relatively linear growth in the time it takes to stabilize a population. In other words, doubling the population size did not double the number of epochs it takes to stabilize the swarm. In fact, in a collinear scheme, the swarm size seemed to correlate less with the stability as some larger swarms outperformed some smaller swarms on average.

\renewcommand{\arraystretch}{1.5}
\begin{table}[H]
\caption{Collinear Initialization Results}

\centering
\resizebox{\columnwidth}{!}{
\begin{tabular}{|c|c||c|c|c|c||c|c|c|c|c|}
\cline{3-11}
\multicolumn{1}{}{} & \multicolumn{1}{}{} & \multicolumn{9}{|c|}{Population Size}\\
\hline
$d:c$ & $r$ & 4 & 8 & 16 & 32 & 5 & 10 & 15 & 20 & 25\\
\hline
\multirow{2}{*}{3:4} & 0 & 25.20 & 181.00 & 0 & 0 & 32.80 & 556.00 & 0 & 0 & 0\\
& 20 & 21.20 & 22.00 & 23.60 & 23.60 & 19.20 & 22.60 & 19.00 & 26.20 & 25.60\\
\hline
\multirow{2}{*}{1:1} & 0 & 0 & 0 & 0 & 0 & 0 & 0 & 0 & 0 & 0\\
& 20 & 27.60 & 57.20 & 88.80 & 88.00 & 43.20 & 55.80 & 76.80 & 73.00 & 99.40\\
\hline
\end{tabular}
}
\vspace{2mm}
\label{rlinearInit}
\end{table}

\subsection*{Capacity}
\subsubsection*{Random}
Because the ideal distance from the target is defined by a static $\delta$ and the range for the distance is defined by a static $\epsilon$, there is a finite stability region for agents to operate in. The area of this region is given by $a$ in Table \ref{capacity_r_Results}. This area determines the number of agents that can reach stability as well as the number of epochs it takes to reach stability. Larger populations take longer to stabilize as more agents try to move closer to the target, yet further from each other within a set region. Because the agents have to get closer to each other to reach stability than in smaller populations, it takes more time to stabilize overall. Some populations are too large for some ideal distances because they simply can't fit all of the agents in the stability region. This can be seen populations of size 128 and 256 when the area of the ideal distance is around 400 square units. Even with a population size of 64, the smallest stability region struggles to carry all of the agents as it takes 210 epochs to reach stability. 

As we can see in the cases with equal area but differing $\epsilon$ and $\delta$, stability is reached quicker when the $\delta$ is larger. Recall that all of the agents are trying to be $\delta$ units from the target. Because the ideal distance is further away, the ideal circle around the target has a larger circumference for agents to tend towards. This allows for stability quicker than smaller $\delta$ values, even when the acceptable area is the same. As expected, smaller $\epsilon$ values take longer to stabilize because it limits the area of the stability region. Further, $\epsilon$ values have less of an effect on experiments with larger $\delta$ values since those values guarantee larger stability regions.

\renewcommand{\arraystretch}{1.5}
\begin{table}[H]
\caption{Random Population Capacity Results}

\centering
\resizebox{\columnwidth}{!}{
\begin{tabular}{|c|c|c||c c c c c c c c c|}
\cline{4-12}
\multicolumn{1}{}{} & \multicolumn{1}{}{} & \multicolumn{1}{}{} & \multicolumn{9}{|c|}{Population Size} \\
\hline
$\epsilon$ & $\delta$ & $a$ & 2 & 3 & 4 & 8 & 16 & 32 & 64 & 128 & 256 \\
\hline
\multirow{3}{*}{2} & 8 & 201.06 & 34.2 & 32.0 & 35.8 & 45.8 & 51.6 & 50.0 & 210.0 & 0  & 0 \\
 & 12 & 301.59 & 29.8 & 21.6 & 31.4 & 44.0 & 37.8 & 40.4 & 66.8 & 0 & 0 \\
 & 16 & 402.12 & 21.2 & 17.6 & 24.2 & 28.0 & 31.6 & 29.6 & 45.0 & 0 & 0 \\
\hline
\multirow{3}{*}{4} & 8 & 402.12 & 35.0 & 32.0 & 32.4 & 34.4 & 48.6 & 48.8 & 45.8 & 53.4 & 0 \\
 & 12 & 603.19 & 26.0 & 28.6 & 18.0 & 35.4 & 46.8 & 42.4 & 41.8 & 48.6 & 355.6 \\
 & 16 & 804.25 & 12.4 & 14.4 & 22.2 & 30.2 & 35.0 & 34.2 & 38.4 & 41.8 & 95.8 \\
\hline
\multirow{3}{*}{8} & 8 & 804.25 & 27.2 & 24.4 & 25.6 & 26.0 & 43.0 & 41.2 & 37.4 & 44.8 & 49.4 \\
 & 12 & 1206.37 & 19.4 & 19.4 & 28.2 & 35.2 & 29.4 & 42.4 & 31.0 &  34.8 & 34.8 \\
 & 16 & 1608.49 &  14.2 & 13.6 & 15.6 & 21.2 & 33.6 & 26.0 & 30.0 & 33.2 & 32.6\\
\hline
\end{tabular}
}
\vspace{2mm}
\label{capacity_r_Results}
\end{table}

\subsubsection*{Boxed}
For the boxed initialization tests for capacity, multiple box positions were again considered. The average of the results from each of these positions (corners, sides, top/bottom, and center) can be seen in Table \ref{capacity_b_Results}. Unlike the random initialization, the boxed initialization values did not align perfectly with the area of the stability range. The number of epochs it took for each swarm to stabilize was minimized by setting larger $\delta$ values. Because all of the agents are aiming toward this $\delta$, the larger it is, the more room there is for agents to move. Since the agents are circling their target, the $\delta$ value is the radius of the circle the agents are following and determines the circumference of that circle. The larger the circumference, the more room for agents to occupy an optimal distance. Because of this, agents are able to stabilize quicker. 

Further, larger $\delta$ and $\epsilon$ values increase the area for agents to swarm and be stable. Therefore, larger $\epsilon$ values also resulted in swarms reaching stability quicker. These values are necessary for larger populations, which have a harder time stabilizing when the area of the stability region is too small. In the most drastic case, it takes a swarm with 64 agents more than 135 epochs to stabilize in a stability region with an area of 200 square units, where it takes the same 64 agents only 17 epochs to stabilize in a region with an area of 1600 square units. Populations larger than this, namely those with 128 and 256 agents, can't be initialized in the boxed regions. Because the boxed regions are set to 10 by 10 units, the only 100 agents can be initialized in a given box. An ideal setup for a boxed initialization would contain less than 64 agents and would have a large $\delta$ value.

\renewcommand{\arraystretch}{1.5}
\begin{table}[H]
\caption{Boxed Population Capacity Results}

\centering
\resizebox{\columnwidth}{!}{
\begin{tabular}{|c|c|c||c c c c c c c c c|}
\cline{4-12}
\multicolumn{1}{}{} & \multicolumn{1}{}{} & \multicolumn{1}{}{} & \multicolumn{9}{|c|}{Population Size} \\
\hline
$\epsilon$ & $\delta$ & $a$ & 2 & 3 & 4 & 8 & 16 & 32 & 64 & 128 & 256 \\
\hline
\multirow{3}{*}{2} & 8 & 201.06 & 27.25 & 22.125 & 23.75 & 25.75 & 22.5 & 28.875 & 136.875 & 0 & 0 \\
 & 12 & 301.59 & 22.625 & 19.25 & 18.125 & 20.375 & 26.5 & 32.75 & 89.625 & 0 & 0 \\
 & 16 & 402.12 & 15.375 & 16.375 & 19.125 & 18.5 & 23.125 & 28.375 & 80.0 & 0 & 0 \\
\hline
\multirow{3}{*}{4} & 8 & 402.12 & 23.75 & 19.375 & 18.75 & 23.25 & 22.25 & 26.625 & 28.625 & 0 & 0 \\
 & 12 & 603.19 & 18.625 & 19.125 & 16.5 & 20.25 & 23.625 & 19.875 & 30.0 & 0 & 0 \\
 & 16 & 804.25 & 14.625 & 12.75 & 14.125 & 18.0 & 18.375 & 21.25 & 26.0 & 0 & 0 \\
\hline
\multirow{3}{*}{8} & 8 & 804.25 & 18.75 & 16.25 & 15.125 & 17.25 & 20.75 & 23.375 & 23.625 & 0 & 0 \\
 & 12 & 1206.37 & 18.0 & 13.5 & 15.25 & 15.375 & 15.0 & 21.75 & 21.625 & 0 & 0 \\
 & 16 & 1608.49 & 9.625 & 8.875 & 14.125 & 13.625 & 14.375 & 17.375 & 16.75 & 0 & 0 \\
\hline
\end{tabular}
}
\vspace{2mm}
\label{capacity_b_Results}
\end{table}

\subsubsection*{Linear}
Due to the size of the field, the maximum population size that can be initialized linearly is 35. Population sizes larger than this have no hope to reach stability because they can't even be initialized. As with the random initialization, we can see that experiments with larger populations and smaller stability areas take longer to stabilize. Again, because the agents need room to move towards the target while moving away from one another, larger stability regions garner faster swarm stability. Further, larger $\delta$ values ensure better stability results for all population sizes. This can again be seen in experiments with the same stability area but differing $\delta$ and $\epsilon$. For instance, swarms with and $\epsilon$ values of 2 or 4 and $\delta$ values of 16 or 8 respectively have different stability regions, but these regions have the same area. Even though the agents have the same amount of area to navigate in, swarms with a larger $\delta$ value see better results. In other words, in a linear initialization, it is better to have a narrower stability region further from the target than a wider stability region closer to the target. 

Compared to random initializations, the linear initialization experiments performed better across the rest of the population sizes. Also, linear experiments where less effected by $\delta$ and $\epsilon$ values. In the random initialization, these values caused larger ranges in stability time. The most drastic difference in stability was for a population size of 256 where it took more than 320 extra epochs to stabilize a swarm in a smaller stability region. However, this is less prevalent in smaller population sizes. If you look at experiments with a population size of 32, the difference from the smallest to the largest $\delta$ and $\epsilon$ values is almost the same in a random experiment than in a linear experiment. More precisely, in the random experiment, it took 24 extra epochs, where in the linear experiments it took 23 extra epochs. 

\renewcommand{\arraystretch}{1.5}
\begin{table}[H]
\caption{Linear Population Capacity Results}
\centering
\resizebox{\columnwidth}{!}{
\begin{tabular}{|c|c|c||c c c c c c c c c|}
\cline{4-12}
\multicolumn{1}{}{} & \multicolumn{1}{}{} & \multicolumn{1}{}{} & \multicolumn{9}{|c|}{Population Size} \\
\hline
$\epsilon$ & $\delta$ & $a$ & 2 & 3 & 4 & 8 & 16 & 32 & 64 & 128 & 256\\
\hline
\multirow{3}{*}{2} & 8 & 201.06 & 30.0 & 33.2 & 26.0 & 32.2 & 36.4 & 43.6 & 0 & 0 & 0\\
 & 12 & 301.59 & 17.2 & 24.4 & 26.2 & 31.6 & 30.6 & 35.6 & 0 & 0 & 0\\
 & 16 & 402.12 & 15.2 & 16.4 & 21.6 & 27.0 & 27.4 & 31.2 & 0 & 0 & 0\\
\hline
\multirow{3}{*}{4} & 8 & 402.12 & 23.2 & 37.2 & 31.2 & 28.2 & 35.6 & 36.4 & 0 & 0 & 0\\
 & 12 & 603.19 & 17.4 & 29.4 & 20.8 & 31.8 & 28.6 & 25.2 & 0 & 0 & 0\\
 & 16 & 804.25 & 19.6 & 22.8 & 11.2 & 26.2 & 21.4 & 29.0 & 0 & 0 & 0\\
\hline
\multirow{3}{*}{8} & 8 & 804.25 & 22.2 & 25.2 & 24.0 & 30.0 & 29.6 & 29.8 & 0 & 0 & 0\\
 & 12 & 1206.37 & 15.6 & 18.0 & 16.0 & 20.8 & 21.6 & 23.8 & 0 & 0 & 0\\
 & 16 & 1608.49 & 11.4 & 11.0 & 15.2 & 14.6 & 19.0 & 20.6 & 0 & 0 & 0\\
\hline
\end{tabular}
}
\vspace{2mm}
\label{capacity_l_Results}
\end{table}

\subsubsection*{Collinear}
As with a linear initialization, the maximum population size that can be initialized collinearly is also 35. This again means that populations above 35 are not capable of being initialized, much less stabilizing. Due to the better performance in linear and collinear initializations over random initialization, they may be better suited for larger population. This indicates that larger fields allowing for larger population sizes would allow for faster stabilization with these initialization types. Further, collinear initialization even beat out some cases of boxed initializations. However, extending the carrying capacity of a box initialization is much easier to accomplish than extending the carrying capacity of a collinear or linear initialization. To extend the capacity for a box setup, one would simply need to increase the size of the box. However, to extend the capacity for a linear or collinear setup, one would need to increase the size of the field itself.

Collinear initializations performed better than both random and linear initializations across all population sizes and distance settings. The best result was with two agents with a wide ideal distance from the target and a large range of acceptable distances. In this case, agents stabilized, on average, in half the time as they did with a random initialization. In another case (population of 32 with $\delta = 8$ and $\epsilon = 2$) agents stabilized, on average, almost twice as fast in a collinear initialization than they did when they were initialized randomly. 

\renewcommand{\arraystretch}{1.5}
\begin{table}[H]
\caption{Collinear Population Capacity Results}
\centering
\resizebox{\columnwidth}{!}{
\begin{tabular}{|c|c|c||c c c c c c c c c|}
\cline{4-12}
\multicolumn{1}{}{} & \multicolumn{1}{}{} & \multicolumn{1}{}{} & \multicolumn{9}{|c|}{Population Size} \\
\hline
$\epsilon$ & $\delta$ & $a$ & 2 & 3 & 4 & 8 & 16 & 32 & 62 & 128 & 256\\
\hline
\multirow{3}{*}{2} & 8 & 201.06 & 22.4 & 24.2 & 25.4 & 33.2 & 33.4 & 34.4 & 0 & 0 & 0\\
 & 12 & 301.59 & 23.4 & 17.4 & 20.2 & 24.4 & 27.4 & 27.2 & 0 & 0 & 0\\
 & 16 & 402.12 & 12.4 & 14.8 & 15.6 & 21.8 & 19.6 & 21.6 & 0 & 0 & 0\\
\hline
\multirow{3}{*}{4} & 8 & 402.12 & 20.6 & 25.0 & 20.4 & 25.6 & 33.6 & 25.0 & 0 & 0 & 0\\
 & 12 & 603.19 & 10.2 & 13.8 & 17.2 & 18.0 & 22.0 & 25.8 & 0 & 0 & 0\\
 & 16 & 804.25 & 6.0 & 16.2 & 13.0 & 19.8 & 18.0 & 19.0 & 0 & 0 & 0\\
\hline
\multirow{3}{*}{8} & 8 & 804.25 & 18.4 & 12.0 & 25.6 & 22.0 & 23.0 & 26.8 & 0 & 0 & 0\\
 & 12 & 1206.37 & 7.4 & 13.0 & 9.4 & 16.0 & 13.8 & 16.4 & 0 & 0 & 0\\
 & 16 & 1608.49 & 7.0 & 10.0 & 7.8 & 10.8 & 13.8 & 11.8 & 0 & 0 & 0\\
\hline
\end{tabular}
}
\vspace{2mm}
\label{capacity_rl_Results}
\end{table}

\section{Conclusion}

\subsection*{Explanation of Results}
The efficiency and capacity of the swarm relies heavily on a number of variables, such as population size, desired distance from the target, acceptable range from the preferred distance, the area of the stability region, the amount of rotation employed by each agent, and the ratio governing the number of units an agent will move to satisfy the Center Rule versus the Dispersion Rule. 

Regarding efficiency, smaller populations with ideal distances further from the target stabilized in fewer epochs than swarms with larger populations or with ideal distances closer to the target. Boxed initializations faired the best, particularly when initialized in the center of the field close to the target. Linear initializations were comparable to the boxed setups, however they were more reliant on a $d:c$ ratio of 3:4. Random initializations performed worse than linear setups, but were more stable than collinear setups which required at least one degree of rotation to reach stability at all.  

Regarding capacity, smaller populations again performed better, especially in experiments with smaller stability regions. Although the area of the stability region was a decent indicator of the performance of a swarm (with the exception of the boxed experiments), a larger preferred distance from the target saw a better outcome in swarms with the same stability region area. Random populations faired the best in capacity results, as the other setups had a population limit on initialization. Linear and collinear populations were limited to the size of the field and boxed populations were limited to the size of the box. These limitations aside, collinear initializations saw the best capacity results. Linear and boxed experiments saw similar results depending on the size of the population. Random populations, while being able to hold more agents in general, took longer to stabilize. 

The results of these experiments show that there are configurations suitable for many real-world applications. Situations that require swarm technology vary greatly. However, the ideal setup can be determined based off of the needs of the user and the results of these experiments. 

\subsection*{Future Work}
Future work for the SHARKS protocol involves dynamic targets. While circling a stationary target is useful for a house fire, it may not be as useful in other situations. Allowing for the target to move around a landscape and still ensuring the circling behavior would extend the applications of this research. Further, the SHARKS protocol can be extended to three dimensional cases. Rather than circling a target, the drones could surround a target, either on the ground or in the air. 

\subsection*{Smart Cities}
These experiments show the efficiency and capacity of the SHARKS protocol. Through looking at different initial agent distributions, population sizes, and ideal distances, the SHARKS protocol allows for agents to circle a target by following only two simple rules. As each target tries to move towards the ideal distance from the target and away from their nearest neighbor, the swarm exhibits a circling behavior around the target. Further, each agent is able to avoid colliding with other agents in the swarm, allowing for safer mobilization of a swarm. Additionally, the protocol allows for the unlikely even that you may lose an agent. The agent itself isn't a threat due to the minimal knowledge it needs of it's surroundings and the loss of the agent doesn't hinder the actions of the rest of the swarm. This furthers research on robotic swarms necessitated by smart cities. Smart cities rely this technology as much as they rely on it to be secure, thereby motivating the decentralized secure-by-design approach of the SHARKS protocol. The protocol does not select lead agents and it is impervious to security threats that some robotic swarms are currently facing. 

\bibliographystyle{IEEEtran}
\bibliography{refs}

\vspace{12pt}

\end{document}